\documentclass{ws-rv9x6arXiv}

\graphicspath{{./figures/evaporation/}{./figures/correlation/}{./figures/apparatus/}
{./figures/outlook/}{./figures/universality/}{./figures/dynamics/}{./figures/sfmott/}}
\usepackage{ws-rv-van}             
\usepackage{epstopdf}

\begin{document}

\chapter{In situ imaging of atomic quantum gases}

\author[Chen-Lung Hung and Cheng Chin]{Chen-Lung Hung$^1$ and Cheng Chin$^2$}

\address{$^1$Norman Bridge Laboratory of Physics 12-33, California Institute of Technology, Pasadena, CA 91125, USA\\
$^2$James Franck institute, Enrico Fermi institute and Department of Physics, University of Chicago, Chicago, IL 60637, USA}

\begin{abstract}
One exciting progress in recent cold atom experiments is the development of high resolution, in situ imaging techniques for atomic quantum gases \cite{Gemelke09, Bakr10, Sherson10}. These new powerful tools provide detailed information on the distribution of atoms in a trap with resolution approaching the level of single atom and even single lattice site, and complement the well developed time-of-flight method that probes the system in momentum space. In a condensed matter analogy, this technique is equivalent to locating electrons of a material in a snap shot. In situ imaging has offered a new powerful tool to study atomic gases and inspired many new research directions and ideas. In this chapter, we will describe the experimental setup of in situ absorption imaging, observables that can be extracted from the images, and new physics that can be explored with this technique.

\end{abstract}

\body

\section{Introduction\label{intro}}

In quantum gas experiments, information about the sample is typically extracted from the optical image of the sample illuminated by lasers near the atomic resonance \cite{ketterle99}. One of the most commonly adopted schemes is the time-of-flight imaging, in which the atoms are first released into free space and then irradiated by the laser after a sufficiently long expansion period. The free expansion is an essential step to reveal the momentum distribution of the sample and is thus very useful for Bose-Einstein condensation experiments, in which many atoms accumulate in the lowest momentum state. It is, however, difficult for the time-of-flight method to uncover the real space information of a trapped sample since the free expansion process easily mixes signal of atoms from different parts of the trap.

\begin{figure}[t]
\begin{center}
\includegraphics[width=0.55\columnwidth,keepaspectratio]{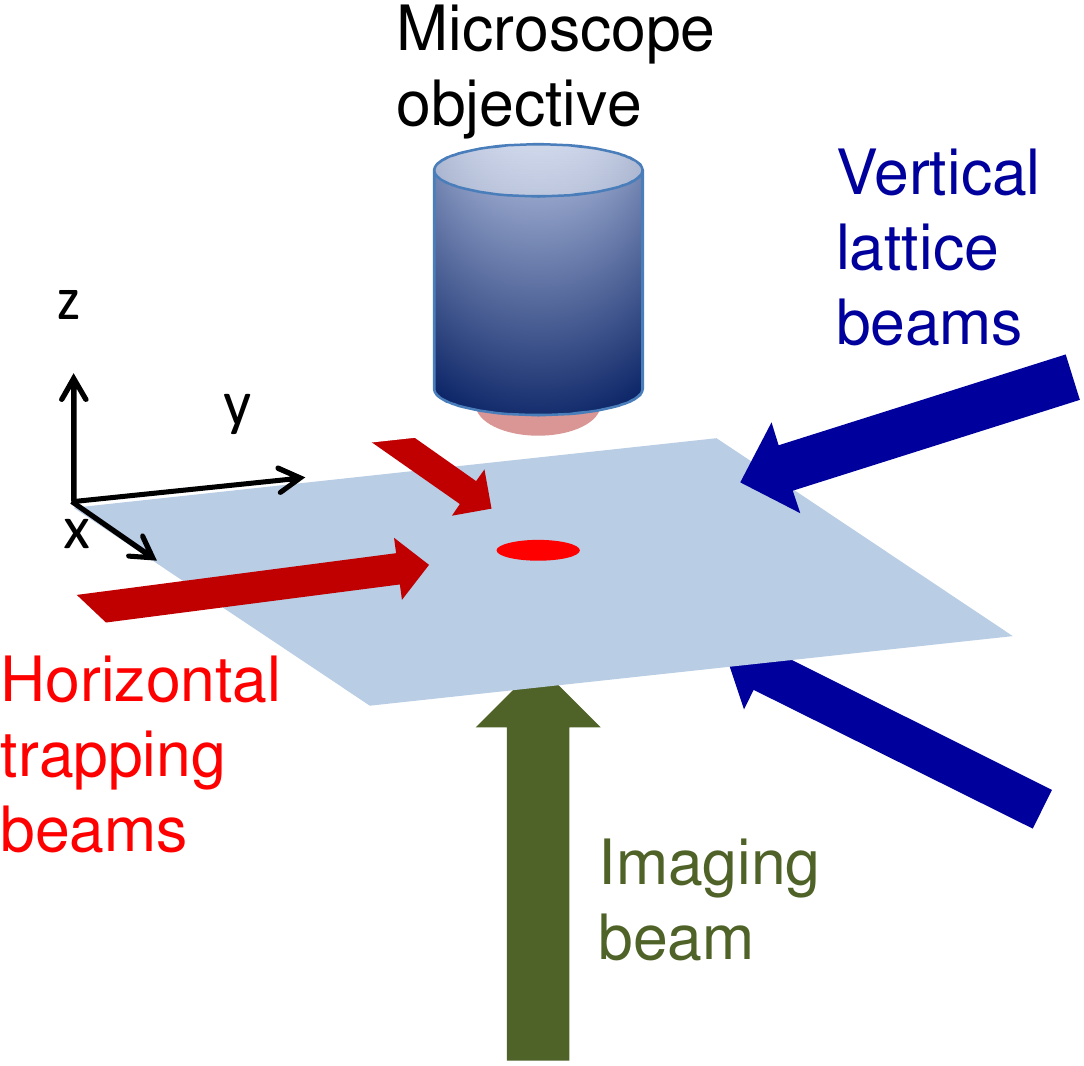}
\caption{In situ imaging of a two-dimensional atomic gas. An atomic sample (red circle) prepared in a 2D optical trap with strong vertical ($z$-direction) confinement provided by the vertical lattice beams, inclined at a small angle relative to the horizontal plane. The atomic sample is loaded into a single lattice site of the lattice in $z$-direction. Horizontal trapping beams provide the radial ($x$- and $y$- directions) confinement. Absorption imaging is implemented by illuminating the sample with an imaging beam in the $z$-direction. Imaging optics and a camera (not shown) are prepared above the sample. } \label{scheme}
\end{center}
\end{figure}

In situ imaging was developed exactly to provide the real space information of the trapped atoms, which not only complements the time-of-flight method, but also turns the density inhomogeneity of the trapped samples into an advantage: a spatially resolved image of a sample reflects its response to different external potential energies. A prominent example is the emergence of a Mott insulator phase in a Bose superfluid in optical lattices, which can be seen from the formation of the density plateau within the sample \cite{Gemelke09, Bakr10, Sherson10}. Furthermore, a careful study of the spatial and temporal variations of the atomic density can help extracting new observables of the system, including the equation of state \cite{Hung11, Yefsah11, Zhang12, Ha13}, density fluctuations \cite{Bakr10, Hung11, Hung11Corr}, and the density-density correlation function \cite{Bakr10, Hung11Corr, Endres11, Cheneau12}.

Compared to the time-of-flight method, in situ imaging method also comes with new technical challenges. The main difficulties include the strong non-linear optical response to dense quantum gases, and the demanding high numerical aperture needed to reach high spatial resolution. To reduce the optical density and to accommodate the very shallow imaging depth of focus, several schemes have been developed to prepare single layer of two-dimensional samples perpendicular to the imaging direction \cite{Gemelke09, Bakr10, Sherson10, Rath10}. Even in this case, as we will discuss below, strong non-linear radiative response can complicate the density calibration. An extreme case is two or more atoms in a tightly confining optical lattice site, where pairs of atoms are lost during the imaging process and the detection yields the parity of the occupation number \cite{Bakr10, Sherson10}. To increase spatial resolution, various designs to implement the \emph{quantum gas microscopy} in optical lattices have been devised and will be discussed in Chapter 7. 

In this chapter, we will describe the experimental approach to performing in situ absorption imaging in Sec.~\ref{setup}, which includes both the preparation of the sample, the construction and optimization of the in situ imaging system. In Sec.~\ref{observables}, we will outline the procedure to extract density, fluctuations and correlations from the images, and examples of intriguing physics that can be explored from these observables.

Given the limited space of the chapter and the fast experimental development, this chapter is sadly unable to address all research efforts in this active field. We will instead provide a more focused and detailed discussions based on selected experiments: the 2D gas and 2D lattice experiments in the Chin group at the University of Chicago and the similar experiments in the Dalibard group at \'{E}cole Normale Sup\'{e}rieure (ENS), Paris. Both experiments perform absorptive in situ imaging.

\section{Experimental setup\label{setup}}
Absorption imaging is commonly employed in time-of-flight measurements, including those that led to the observation of the first atomic Bose-Einstein condensation in 1995. In these experiments, absorption imaging proceeds by illuminating an expanding sample with a short pulse of laser beam, and then recording the shadow pattern cast by the atoms. From the known absorption cross section of an atom, the density distribution of the expanded sample is determined from the shadow.

In both the ENS and the Chicago groups, in situ absorption imaging is performed on atomic gases localized in a trap with very strong axial ($z$-direction) and weak radial ($x$- and $y$- directions) confinements, and the atomic density distribution in the radial direction is extracted. The trap forces almost all atoms to populate the motional ground state in the axial direction, effectively compressing the sample into a two-dimensional (2D) one. This also greatly reduces the optical density in the imaging direction. In the radial directions, the sample is typically confined by a weak harmonic dipole trap or by a 2D optical lattice. In the following discussions, we call the sample a 2D gas in the former case, and a 2D lattice gas in the latter case. One scheme for in situ absorption imaging of a 2D gas is shown in Fig.~\ref{scheme}.

\subsection{Preparation of a two-dimensional sample}\label{abs:prepare}
The preparation of a single layer of 2D sample begins by compressing a regular Bose-Einstein condensate (BEC) into a tight trap. The ENS group uses an elliptic blue-detuned beam with a single nodal plane, and the Chicago group uses a single site of a one-dimensional vertical lattice, formed by overlapping two red-detuned laser beams with a small intersecting angle; see Fig.~\ref{scheme}. The intensity profile of the trapping beam adds only a weak horizontal confinement to the sample.

When the lattice approach is adopted, care should be taken to ensure all atoms are loaded into a single site. In the Chicago experiment, cesium atoms are first precooled and condensed into an oblate dipole trap whose short axis is aligned in the vertical direction. The trap is formed by laser beams propagating in the horizontal $x$-$y$ plane (Fig.~\ref{scheme}). The trap frequencies are $(\omega_x, \omega_y, \omega_z)=2\pi\times (12, 17, 60)$~Hz, and the vertical Thomas-Fermi radius of the condensate 1.5$~\mu$m is small compared to the vertical lattice spacing of 3.1~$\mu$m. The oblate dipole trap is carefully aligned to match one site of the vertical lattice, and the loading is implemented by slowly ramping on the vertical lattice in the $z$-direction for 800~ms. The alignment is critical and may drift over time, but a fidelity of $>98\%$ into a single site is constantly reached after minor tweaking.

The single site loading efficiency is confirmed based on microwave tomography. In the presence of a vertical magnetic field gradient, atomic populations in different lattice sites are distinguished by driving the atoms to a different hyperfine state via a field sensitive microwave transition and imaging only the atoms in the new hyperfine state. The quoted fidelity is reached by fine tuning the vertical position of the condensate before the loading process. At the end of the day, 4 to $5\times 10^4$ atoms are typically prepared in a single site with the trap frequencies $\mathbf{\omega }=2\pi\times (8, 10, 2000)$~Hz.
	
In addition to loading atoms into a single lattice site, a greater challenge is to reduce the imperfection of the dipole trap potential, which can be easily seen from the condensate density profiles. Dipole trap potential corrugations can come from diffraction and interference of the trapping beams. Position drifts can be caused by thermal contraction of the optical modulators. It is thus highly advisable to couple the laser beams into fibers to clean up the mode and put optics and modulators before the fiber coupling if possible. Furthermore, feedback control and stabilization of the beam intensity is essential to obtain a long lifetime of the sample and the reproducibility of the measurement.

\begin{figure}[t]
\begin{center}
\includegraphics[width=0.275\columnwidth,keepaspectratio]{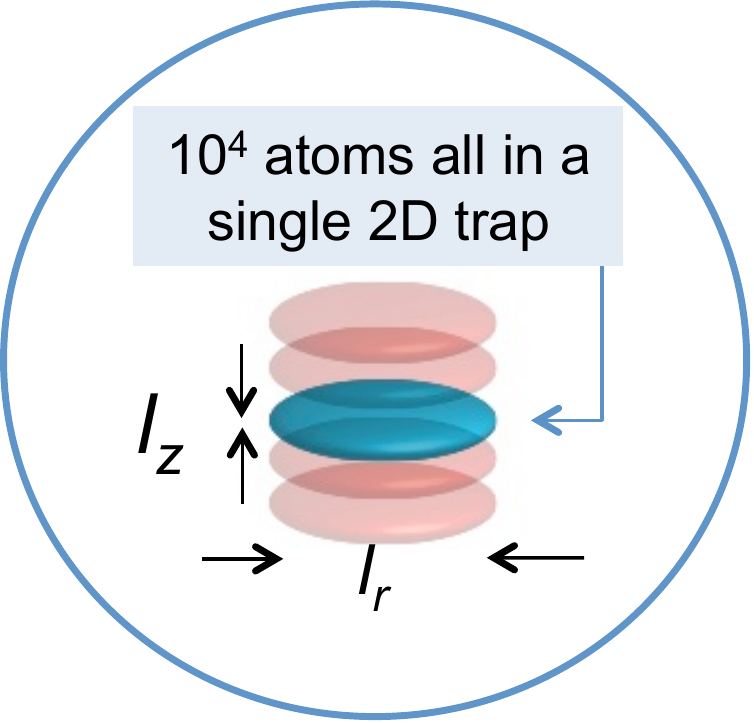}
\includegraphics[width=0.45\columnwidth,keepaspectratio]{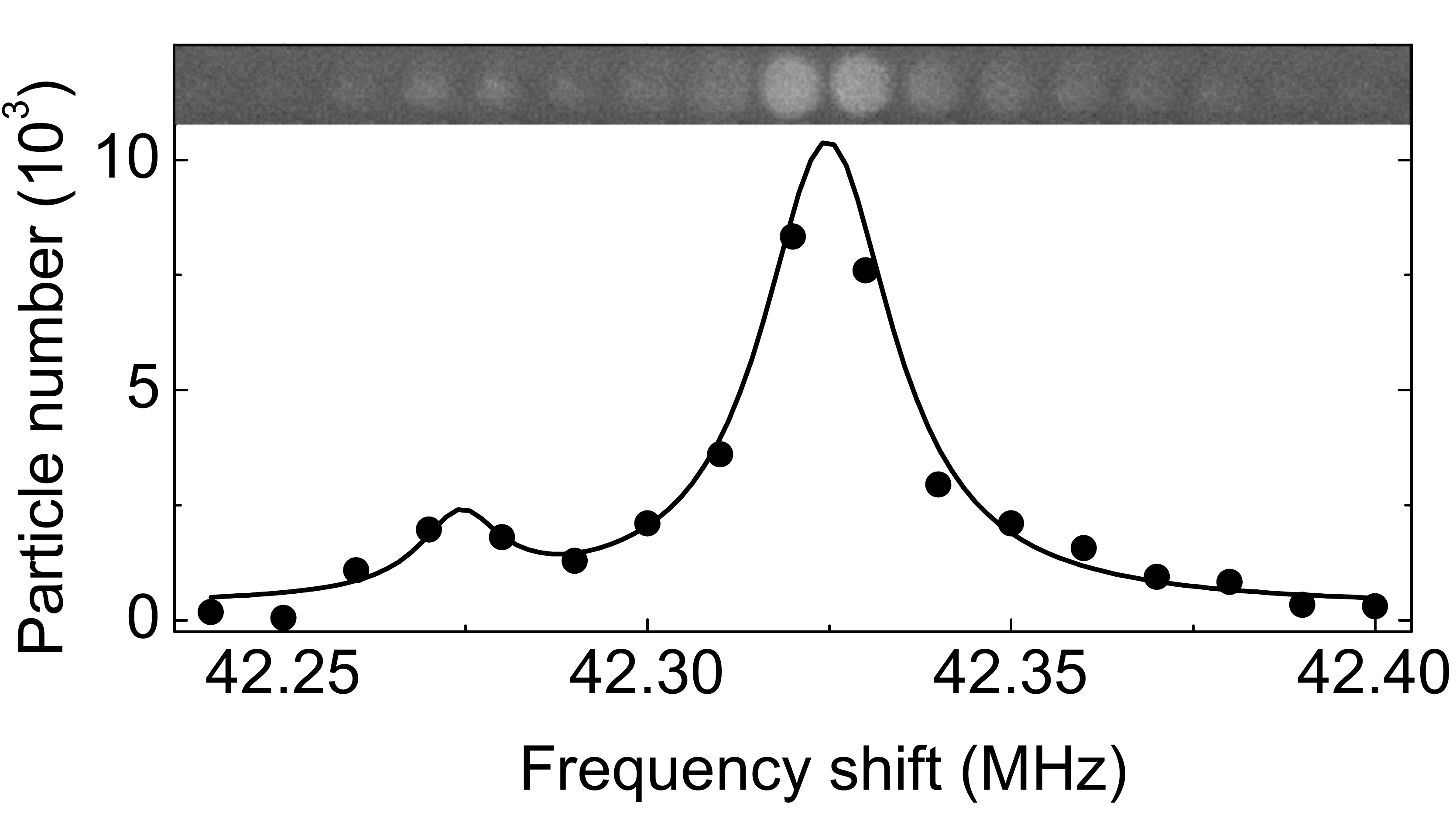}
\caption{Microwave tomography to distinguish atoms residing in different lattice sites. By applying a magnetic field gradient along the lattice direction, atoms in different sites can be selectively driven by microwaves and imaged. Here, the main peak at 42.325~Mhz corresponds to the population in the desired lattice site, and the side peak at 42.275~Mhz the small population in the neighboring site. Upper panel shows the images at different driving frequencies. The microwave tomography is used to optimize the single site loading with $>98\%$ efficiency \cite{Hung11thesis}. Vertical (horizontal) trap harmonic oscillator length is $l_z = 195$~nm $(l_r=2.9~\mu$m).} \label{tomography}
\end{center}
\end{figure}

For 2D lattice experiments in the Chicago group, additional two retro-reflected beams of the two horizontal trapping beams are turned on to form a 2D square lattice potential with a lattice constant of $\lambda_L/2=0.532~\mu$m, where $\lambda_L$ is the lattice beam wavelength. The two retro-reflected beams are independently controlled via a pair of acousto-optic modulators, which offer both high modulation bandwidth of $>1$MHz and large dynamic range $\sim$50~dB. The retro beams are slowly turned on in 300~ms to minimize the heating due to non-adiabatic processes \cite{Hung10}.

\subsection{Imaging system \label{abs:image}}
The in situ absorption imaging is performed in the same way as conventional absorption imaging. The sample is illuminated with a resonant imaging beam and the transmitted intensity distribution is recorded by a charge-coupled device (CCD) camera. Microscope objectives or optics with high numerical aperture (N.A.) and low aberrations are preferred to obtain high imaging resolutions. A diffraction limited imaging system should offer optical resolution given by $R=0.61 \lambda/$N.A., where the resolution $R$ is defined based on the Rayleigh criterion and $\lambda$ is the imaging beam wavelength.

\begin{figure}[t]
\begin{center}
\includegraphics[width=.8\columnwidth,keepaspectratio]{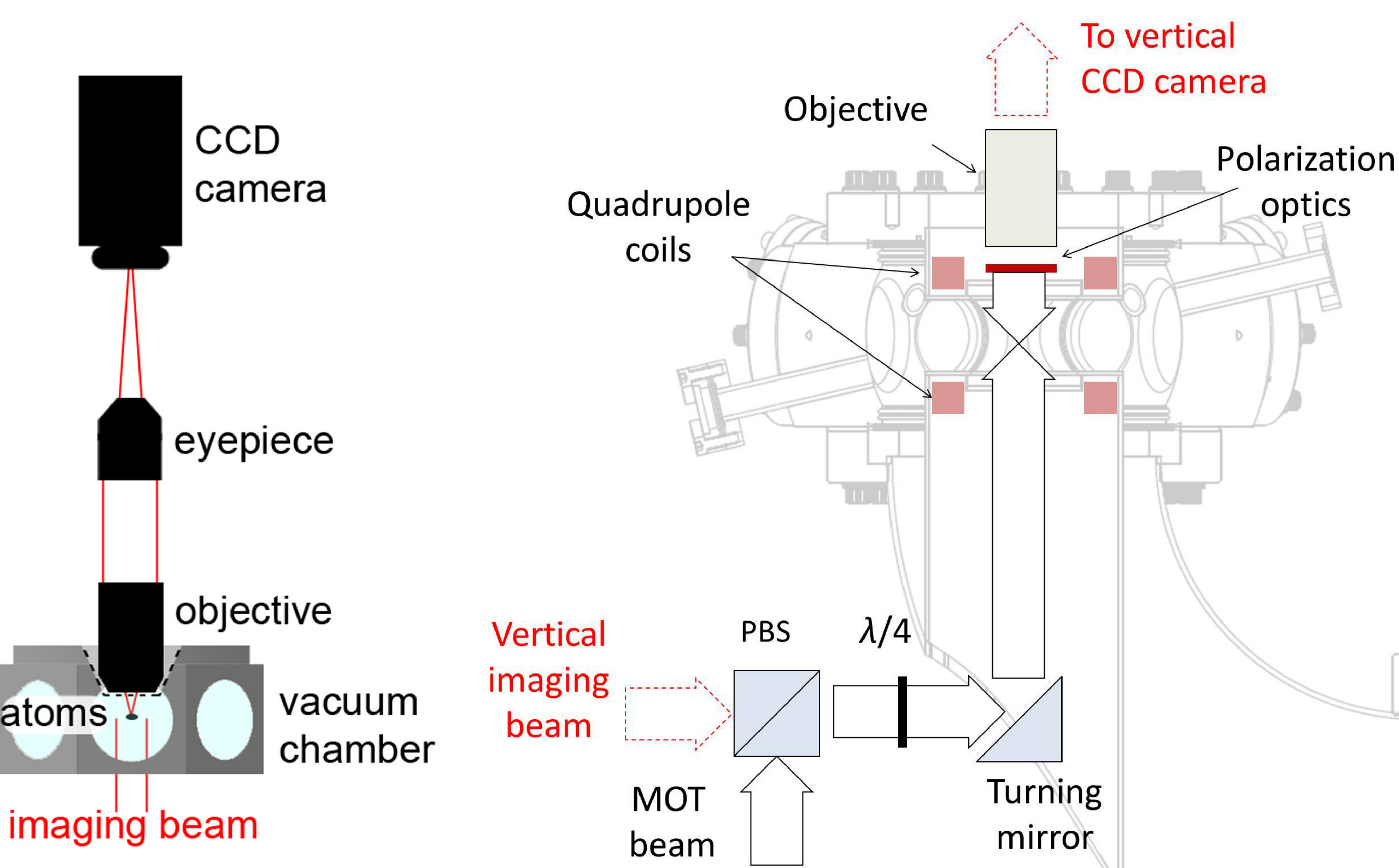}
\caption{Setup of in situ imaging of 2D atomic samples in the Chicago group. Left panel: An imaging beam is sent from below the chamber and its transmission captured by the objective, eyepiece and the camera. Right panel: Details of the setup near the chamber illustrate the combination of MOT and imaging beams in the same beam path. Here the polarization optics (red line) right above the viewport include a quarter-wave plate and a reflective polarizer that reflects the MOT beam and transmits the imaging beam. With this setup, at a working distance of 34~mm and N.A.$=0.28$, a resolution of 1.8~$\mu$m is reached with a commercial objective. Based on a custom objective (Special Optics) at a working distance of 24~mm and N.A.$=0.5$, a diffraction-limited resolution of $1.0~\mu$m is reached.}\label{imaging_system}
\end{center}
\end{figure}

A convenient and inexpensive choice to implement in situ imaging is based on a commercial infinity-corrected microscope objective, a tube lens and a CCD camera. Care should be taken to select the objective to maximize the N.A. given the available optical access, to minimize etaloning, aberrations and misalignment, and to ensure the mechanical stability. At a working distance of 30$\sim$40~mm, available to typical experiments, a resolution of 2 microns or better can be reached with standard commercial objectives that cost $<2$k USD.

\subsection{Optimization of in situ imaging \label{abs:optimization}}

Performance of high resolution optical microscopy depends on the optical access, quality and alignment of the optics. A general guideline is to design the imaging system that preserves the numerical aperture available to the system, with low aberrations and proper optical coating. A standard bench test that faithfully simulates the real imaging setup is valuable to determine the resolution, as well as to characterize the depth of focus and the field of view.

When we move on to imaging cold atoms, a common empirical approach to optimize the optical alignment is to progressively image smaller samples or to distinguish finer features in the density profile. As an example, Bose condensates with the scattering length tuned to small values can be simple and calibrated targets to optimize imaging. Another example is atoms in optical lattices formed by beams with small intersecting angles $\theta \ll 1$ radian and thus a large lattice spacing of $\lambda_L/(2\sin\theta/2)$.

Due to the limited depth of focus, it is essential to minimize the displacement of the atoms during imaging, and thus a short ($\sim$10~$\mu$s) pulse of the imaging beam with intensity $I>I_s$ higher than the saturation intensity $I_s$ of atom-photon scattering is frequently desired. This is in contrast to conventional absorption imaging of low density gases which typically operates at $I \ll I_s$. The short and intense pulse is intended to extract as many scattered photons ($\sim$300) per atom as possible before they are pushed out of the depth of focus, therefore maximizing the signal-to-noise ratio. The high imaging beam intensity is also beneficial for the extraction and calibration of the atomic density, see Sec.~\ref{abs:density}.

\subsection{Aberrations and modulation transfer function  $\mathcal{M}(\mathbf{k})$ \label{abs:mtf}}

One of the common limitations in high-resolution imaging is optical aberrations, which can come from either the misalignment, the imperfection of the optical elements or both. In the following, we will first outline the working principle of absorption imaging and then discuss an empirical approach to characterize and reduce imaging aberrations.

We consider a single atom illuminated by a uniform imaging beam. The incident field $E_0$ is scattered by the atom in the object plane, and optically collected and focused in the image plane (CCD camera). The scattered field on the camera can be written as $\Delta E\propto E_0 e^{i\delta_s} p(\mathbf{k})$, which is proportional to the incident field $E_0$, and carries a phase shift $\delta_s$ due to the laser detuning. Here $p(\mathbf{k})$ is the Fourier transform of the exit pupil function of the imaging optics, and $\mathbf{k}=\mathbf{r}/ad$ relates the position $\mathbf{r}$ in the object plane to the position $\mathbf{R}=M\mathbf{r}$ in the image plane. Here $M$ is the imaging magnification, $a$ the limiting aperture, and $d = \lambda / (2 \pi \mathrm{N.A.})$ \cite{Goodma05}. For a perfect optical system, as an example, $p(\mathbf{k})$ is a delta function. The absorption image of an atom, called the point spread function $\mathcal{P}$, is formed by interfering the scattered field with the incident field in the image plane. Given the transmission of the incident beam of $t^2=|E_0+\Delta E|^2/|E_0|^2\approx 1+2\Re[\Delta E/E_0]$, we obtain the relation $\mathcal{P}(\mathbf{r}) \propto \Re [e^{i\delta_s}p(\mathbf{k})]$, where $\Re[.]$ refers to the real part.

To account for aberrations, intensity attenuation and truncation introduced by the optical setup, we can model the exit pupil function as
\begin{eqnarray}
p(r_p,\theta_p)=U(r_p/a,\theta_p)e^{i \Theta(r_p/a,\theta_p)},\label{pupil}
\end{eqnarray}

\noindent where $r_p$ and $\theta_p$ are polar coordinates on the exit pupil, $U(\rho,\theta)$ is the transmittance function, and $\Theta(\rho, \theta)$ is the wavefront aberration function. An empirical model for $U$ is $U(\rho)= H(1-\rho)e^{-\rho^2/\tau^2}$, where the Heaviside step function $H(x)$ sets the sharp cut-off of the aperture and, in the absence of other imperfections, leads to the Airy pattern of the point spread function in fluorescence imaging. The factor $e^{-\rho^2/\tau^2}$ can be introduced to model the weaker transmittance at large incident angle due to, e.g., finite acceptance angle of optical elements.

For high quality optics with low aberrations, the aberration function $\Theta(r_p/a,\theta_p)$ can be approximated by

\begin{equation}
\Theta(\rho,\theta) \approx S_0 \rho^4 + \alpha \rho^2 \cos(2\theta - 2 \phi) + \beta \rho^2 ,\label{waveaber}
\end{equation}
where $S_0$ characterizes the spherical aberration, $\alpha$ the astigmatism (with $\phi$ the azimuthal angle of the misaligned optical axis), and $\beta$ the defocusing due to atoms deviating from the focal plane during the imaging. Further details of the aberration theory in optical imaging can be found in Ref.~\cite{Hung11Corr}.

\begin{figure}[t]
\begin{center}
\includegraphics[width=.25\columnwidth,keepaspectratio]{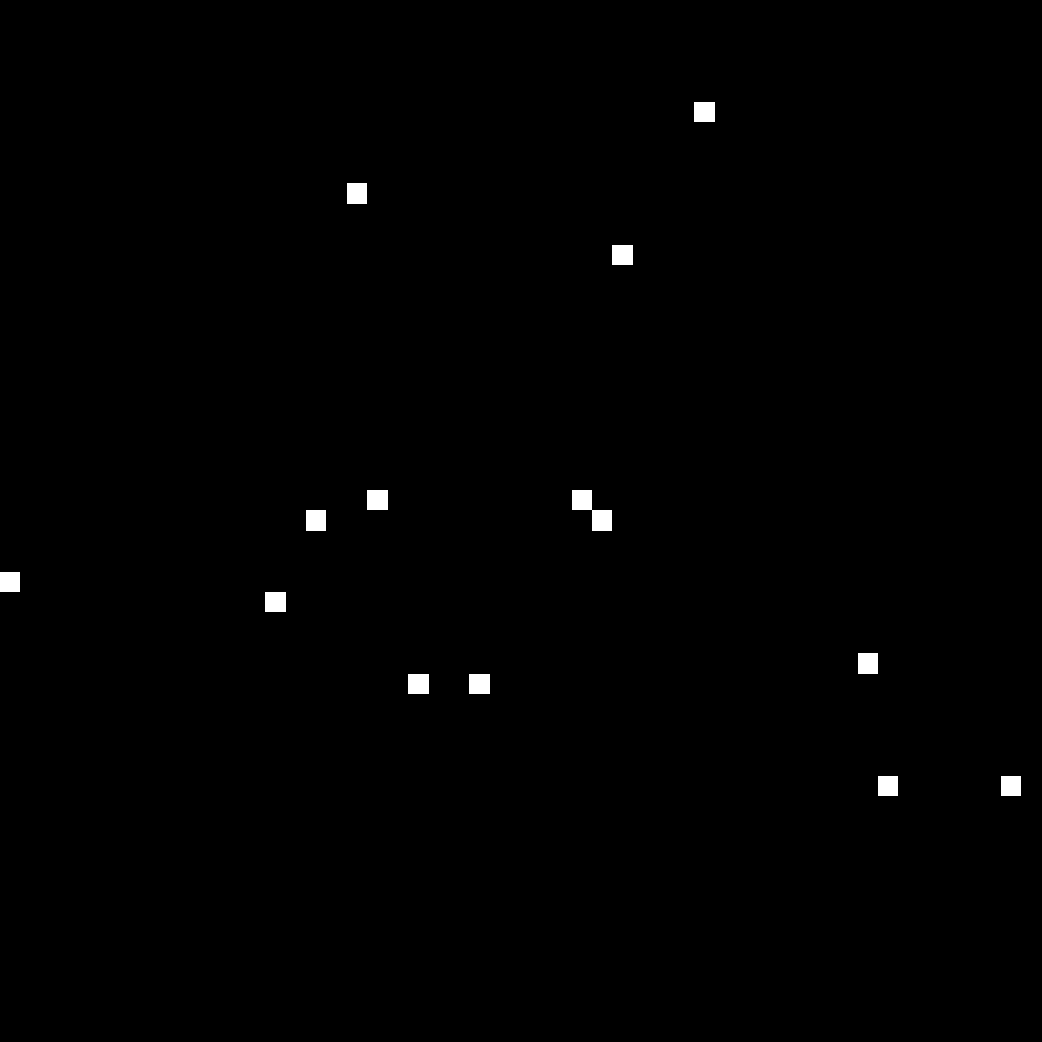}
\includegraphics[width=.25\columnwidth,keepaspectratio]{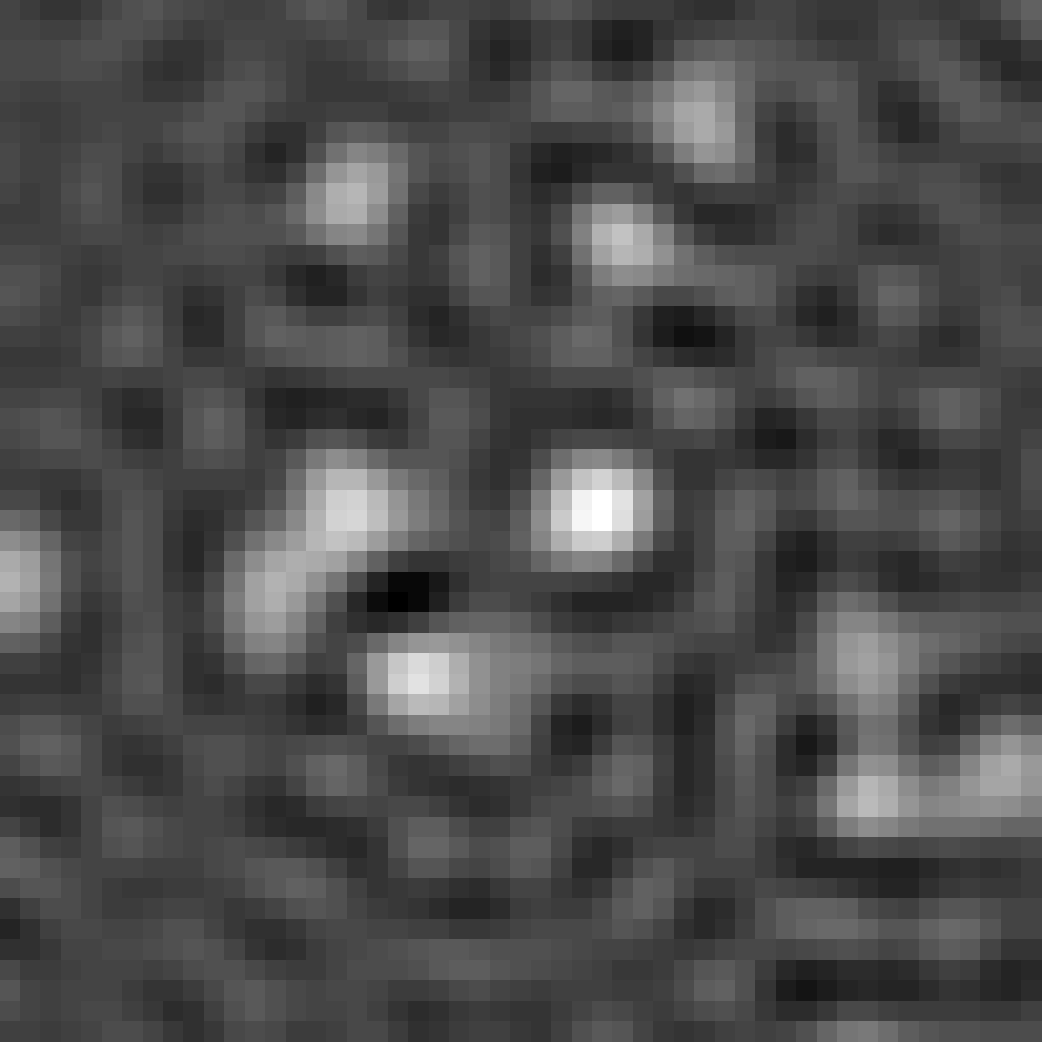}
\caption{Illustration of an atomic distribution and its optical image. (Left panel) A sample of randomly distributed atoms. Each bright dot represents an atom. (Right panel) Optical image of the atoms calculated based on the convolution, described in Eq.~(\ref{nexp}), with a model point spread function.}\label{psf}
\end{center}
\end{figure}

On the experimental side, one may determine the point spread function $\mathcal{P}$ directly if a single atom can be isolated and imaged. Here we outline an alternative method based on a thermal gas, which is accessible to most cold atom experiments. We assume each CCD pixel is smaller than the image resolution in the image plane and atomic displacement during the imaging process is small compared to the pixel size in the object plane. The atom number $N_j$ recorded on the $j$-th CCD pixel and the corresponding density $n_{exp} (\mathbf{r}_j)$ is given by

\begin{equation}
n_{exp} (\mathbf{r}_j) \equiv \frac{N_j}{A} \approx \int n(\mathbf{r}) \mathcal{P}(\mathbf{r}_j-\mathbf{r}) d^2 r,\label{nexp}
\end{equation}

\noindent where $A$ is the area of a CCD pixel, $\mathbf{r}_j$ is the center position of the $j$-th pixel in the object plane, $n(\mathbf{r})$ is the real atomic density distribution, and the integration runs over the entire $x-y$ coordinate space; see Fig.~\ref{psf} for an example. Introducing the density fluctuation $\delta n=n-\bar{n}$ as the deviation of a single shot measurement from the mean of many, we can rewrite the equation in Fourier space as
\begin{equation}
\delta n_{exp} (\mathbf{k}_l)\equiv \sum_j \delta N_j e^{-i \mathbf{k}_l\cdot \mathbf{r}_j}  \approx  \delta n(\mathbf{k}_l) \mathcal{P}(\mathbf{k}_l),\label{d_nexp}
\end{equation}
where $\delta N_j = N_j - \bar{N}_j$, and $\mathbf{k}_l=\frac{2\pi}{L}(l_x,l_y)$ with $l_x$ and $l_y$ being integer indices in $\mathbf{k}$-space and $L$ the linear size of the image.

\begin{figure}[t]
\begin{center}
\includegraphics[width=0.9\columnwidth,keepaspectratio]{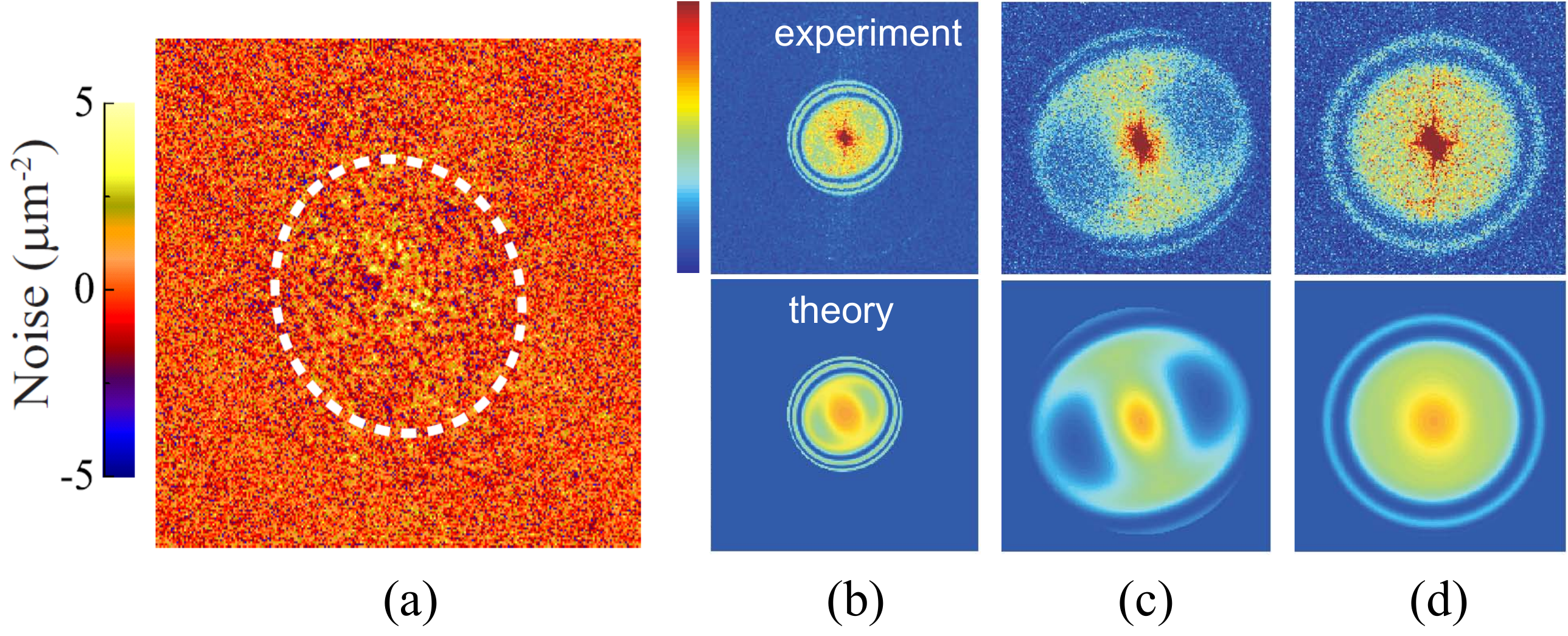}
\caption[Determination of the modulation transfer function $\mathcal{M}(\mathbf{k})$ from \textit{in situ} images of 2D thermal gases.]{Determination of the modulation transfer function $\mathcal{M}(\mathbf{k})$ from \textit{in situ} images of 2D thermal gases. For atomic cesium, a normal gas at high temperature $>100$~nK is uncorrelated down to the length scale of the thermal de Broglie wavelength $\lambda_{dB} < 0.5 \mu$m \cite{Naraschewski99}, which is below our imaging resolution. Therefore, the density fluctuation power spectrum of the sample reveals the square of the modulation transfer function, as indicated by Eq.~(\ref{npower}). A sample image of the fluctuation distribution is shown in (a), obtained by subtracting the averaged density image (60 shots) from a single-shot image, namely, $n(x,y)-\bar{n}(x,y)$. The image size is $256 \times 256$ pixels or $(170\mu$m$)^2$ in the object plane, and the dashed ellipse indicates the location of the atoms. Outside the circle, the fluctuations are dominated by the optical shot noise, which is independent of the spatial frequency. Within the circle, the excess fluctuations come from the atoms. Samples of the 2D fluctuation power spectra $\langle |\delta n_{exp} (\mathbf{k})|^2 \rangle /  N$ are shown in (b), for a commercial objective with N.A.=0.28 and in (c), for a custom objective with N.A.=0.50. Each spectrum has a sharp diffraction-limited edge at $k=2\pi \mathrm{N.A.}/\lambda$. Ripples close to the edge are due to imaging aberrations. The cylindrical asymmetries of the spectra in (b) and (c) are the results of the astigmatism due to a misaligned optical axis. In (d), the spectra are obtained after the removal of the misalignment. The theoretical fits to the experimental data are generated using the imaging response function defined as $\mathcal{M}^2_{fit}=|\mathcal{FT}(\Re[e^{i\delta_s}\mathcal{FT}(p)])|^2$, where $\mathcal{FT}(.)$ denotes the Fourier transform, and the aberration parameters defined in Eqs.~(\ref{pupil}) and (\ref{waveaber}). All panels in (b-d) have the size of $k_x,k_y=-5.9\sim5.9/\mu$m. Figure (a) adapted from Hung et al.\cite{Hung11Corr}} \label{corrfig2}
\end{center}
\end{figure}

For thermal gases at high temperature, fluctuations are uncorrelated and are given by the shot noise $\langle |\delta n(\mathbf{k}_l)|^2 \rangle=N$, where $N$ is the total atom number inside the image. Here $\langle ... \rangle$ denotes ensemble averaging and can be evaluated by averaging over measurements from repeated experiments. Thus we obtain
\begin{equation}
\langle |\delta n_{exp} (\mathbf{k})|^2 \rangle \approx  N \mathcal{M}^2(\mathbf{k}),\label{npower}
\end{equation}
where the modulation transfer function $\mathcal{M}(\mathbf{k})\equiv |\mathcal{P}(\mathbf{k})|$ can be measured experimentally and fit by a theoretical model; see Fig.~\ref{corrfig2}. From the fits, the aberrations can be estimated and used to refine the optical alignment. Determination of the modulation transfer function is also essential for the extraction of the static structure factor, discussed in Sec.~\ref{abs:correlations}.

For simplicity, in the above and the forthcoming discussions, we do not include the contribution of optical shot noise from the imaging beam, which will certainly contribute to the noise of the measured atomic density $\delta n_{exp}$. Nonetheless, for imaging beams with spatially uniform intensity, the optical shot noise is spectrally uniform and contributes to a constant background in $\langle |\delta n_{exp} (\mathbf{k})|^2 \rangle$. This background can be reliably measured at high spatial frequencies beyond the diffraction limit of the imaging system, where $\mathcal{M}^2(\mathbf{k}) = 0$, and can be subtracted off at all $\mathbf{k}$, leaving our analysis unaltered.

\section{Physical observables from in situ images \label{observables}}
In addition to atomic density, density fluctuations and correlations are also major observables from in situ images. Extraction of these quantities from images, however, can suffer from various non-linear effects and experimental systematics that require careful evaluation. We will discuss practical approaches to reduce the systematics and to calibrate the measured density and fluctuations in this section. From the observables, we will also outline a procedure for deriving interesting physical quantities, including the equation of state, the scaling functions and the density-density structure factors.

\subsection{Extraction and calibration of atomic density \label{abs:density}}
While our imaging scheme is similar to those of conventional absorption imaging, radiative interactions between an imaging beam and a 2D atomic sample can be significantly different from those with a three-dimensional sample, even when both samples are prepared to have the same column density along the imaging direction. This disparity comes from the fact that the spatial extent of a 2D sample is typically smaller than the optical wavelength of the imaging beam $\delta z<\lambda$, and photon scattering within the sample suffers from severe multiple scattering and reabsorption processes even with modest optical density. Experimental \cite{Rath10} and numerical \cite{Chomaz12} studies show that a strong nonlinear correction to the radiative cross section exists for 2D samples interacting with a weak imaging beam with intensity $I_0<I_s$, where $I_s$ is the saturation intensity. This is because the 2D geometry leads to a scattered radiation field from other atoms that is comparable in strength with the incident field.

\begin{figure}[t]
\begin{center}
\includegraphics[width=0.75\columnwidth,keepaspectratio]{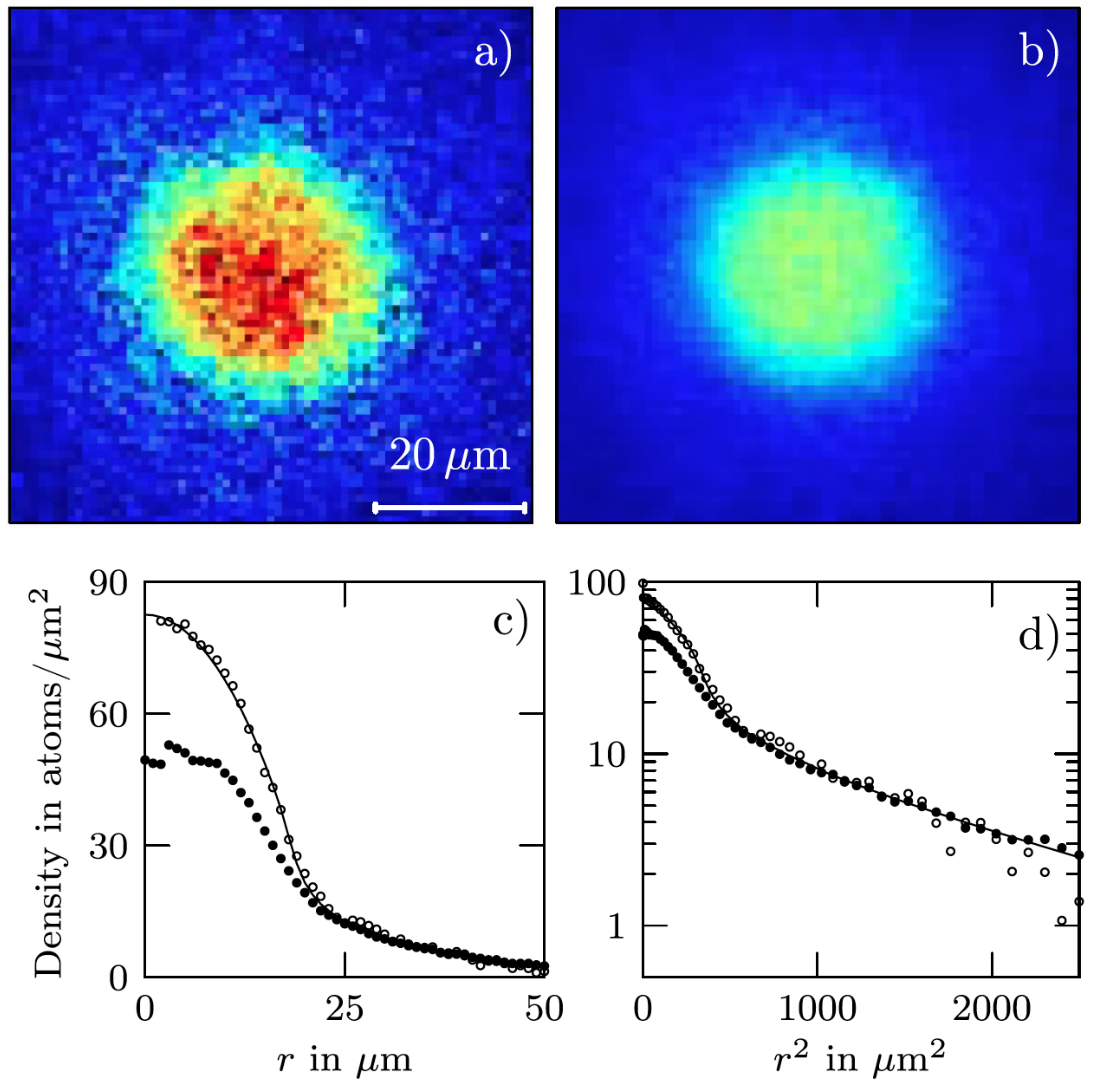}
\end{center}
\caption{Absorption imaging of 2D clouds with (a) strong ($I_0/I_s=40$) and (b) weak ($I_0/I_s=0.5$) probes. The images are converted to atomic density profiles using the modified Beer-Lambert's law Eq.~\ref{modified_beer}. (Lower panels) Comparing the radial density profiles of image (a) (open circle) and image (b) (filled circle) to a mean-field theory calculation (solid line), one obtains better consistency either for the measurement with a strong imaging pulse or at regions with lower atomic density. Figures adapted from Yefsah et al.\cite{Yefsah11}. Copyright (2011) by the American Physical Society.}
\label{fig_strong_abs}
\end{figure}

This nonlinear complication can be suppressed by employing an intense imaging beam with intensity $I_0 \gg I_s$ \cite{Hung11, Yefsah11}; see Fig.~\ref{fig_strong_abs}. Here the strong imaging beam saturates the atoms and thus reduces the influence of the scattered field from other atoms. Satisfying the above criteria makes the interpretation of absorption imaging in a 2D geometry much simpler. In this case, one can derive the atomic density based on the modified Beer-Lambert's law \cite{Reinaudi07, Yefsah11}:

\begin{eqnarray} \label{modified_beer}
n=\frac{1}{\sigma} \Big(\ln\frac{I_0}{I_t}+\frac{I_0-I_t}{I_s} \Big),
\end{eqnarray}

\noindent where $\sigma$ is the radiative cross section of a single atom. Experimentally, $I_s$ can be determined from the intensity dependence of the optical density, while $\sigma$ can be calibrated by the density dependence of the Thomas-Fermi radii for an oblate Bose condensate or from local density fluctuations of an uncorrelated normal gas \cite{Hung10, Hung11thesis}.

A sample 2D atomic density profile derived from in situ imaging using the above approach is shown in the left panel of Fig.~\ref{figEoS}.

\subsection{Local density approximation and equation of state\label{eos}}
 The inhomogeneous trapping potential is often a nuisance in the analysis of the time-of-flight images, but becomes an advantage with in situ images. The spatial profile of the sample reveals its response to different chemical potentials. This association, called the local density approximation (LDA), allows us to extract the equation of state (EoS) $n(\mu,T)$ of the substance from the density distribution of a trapped sample $n(\mathbf{r})$ and the independently-measured chemical potential $\mu$ and temperature $T$.

The local density approximation in cold atom experiments suggests that quantities measured at or near a position $\mathbf{r}$ with potential energy $V(\mathbf{r})$ reflect those in a homogeneous system with a properly chosen chemical potential $\mu$. This approximation is valid for systems in thermal equilibrium confined in a slowly varying external potential $V(\mathbf{r})$, where the gradient of the quantum pressure can be ignored. Here the trapped sample can be viewed as many subsystems at different location $r$, and the equilibrium is established when the thermodynamic force balances the mechanical force $-\nabla \mu(\mathbf{r})-\nabla V(\mathbf{r})=0$. Equivalently, we have $\mu(\mathbf{r})+V(\mathbf{r})=\mu_0$, where the constant $\mu_0$ can be identified as the chemical potential at the trap center with $V(\mathbf{r}=0)\equiv 0$. Under the LDA, any local observable $x(\mathbf{r})$ can thus be associated with a subsystem with $\mu(\mathbf{r})=\mu_0-V(\mathbf{r})$. Thus an in situ measurement allows us to convert a local observable $x(\mathbf{r})$ into $x(\mu)$ of a homogeneous system with widely tunable chemical potentials as $x(\mu)|_{\mu=\mu_0-V(\mathbf{r})}=x(\mathbf{r})$.

\begin{figure}[t]
\begin{center}
\includegraphics[width=0.4\columnwidth,keepaspectratio]{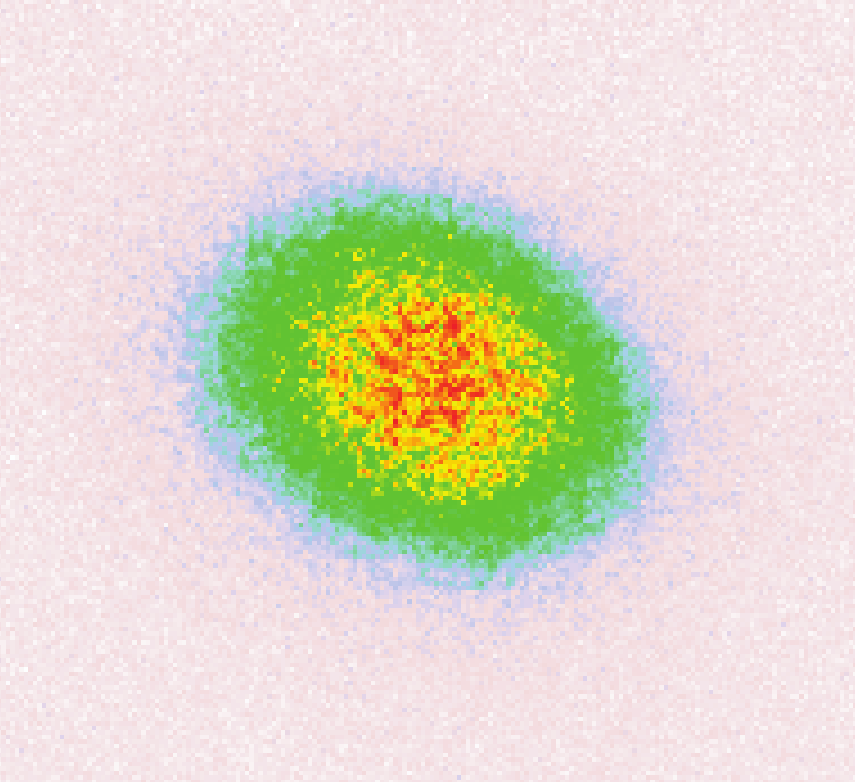}
\includegraphics[width=0.525\columnwidth,keepaspectratio]{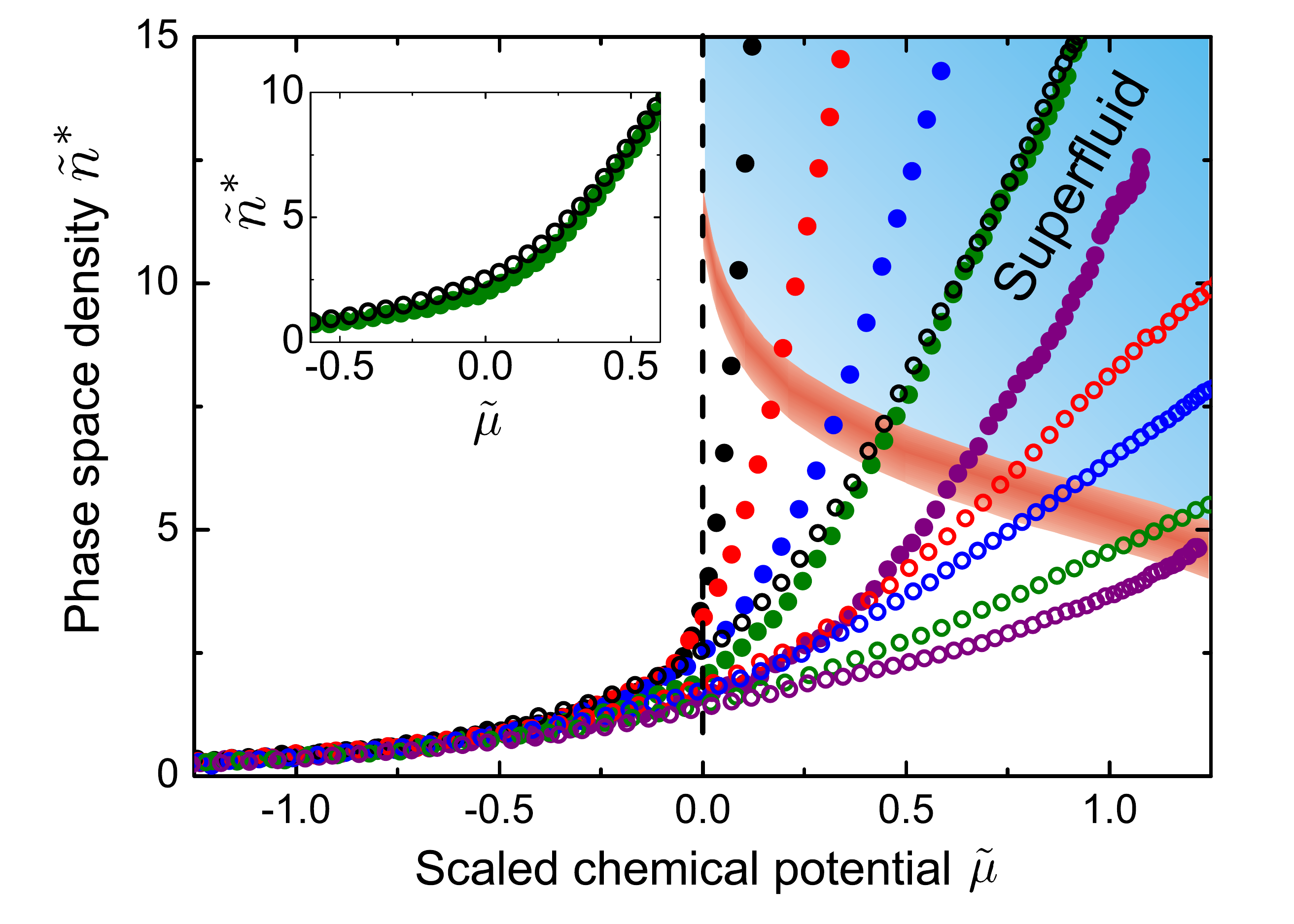}
\end{center}
\caption{Density profile and equations of state for 2D Bose gases and 2D lattice gases. (Left panel) In situ density profile of 4$\times 10^4$ atoms in a 2D trap. Image resolution is 1.0~$\mu$m and each pixel corresponds to (0.533~$\mu$m)$^2$ in the object plane. The image is taken based on an objective with N.A.=0.50 and 24~mm working distance and is averaged over 10 samples. (Right panel) Derived equations of state from the density profiles of samples in thermal equilibrium in the dimensionless units of $\tilde{\mu}=\mu/k_BT$ and phase space density $\tilde{n}^*$ ($*$ indicates the inclusion of effective mass correction for lattice gases). The filled circles refer to 2D gases with coupling constant $g=0.05$ (black), 0.15 (red), 0.24 (blue), 0.41 (green) and 0.66 (purple), controlled by a Feshbach resonance. The open circles are 2D lattice gases with $g=0.45$ (black), 0.85 (red), 1.2 (blue), 1.9 (green) and 2.8 (purple). The blue and red shaded area mark the 2D superfluid regime and the phase boundary, respectively. The inset compares the equations of state of a 2D gas and a 2D lattice gas with an almost identical $g\approx 0.4$. Figures adapted from Ha et al.\cite{Ha13}}\label{figEoS}
\end{figure}

\begin{figure}[h]
\begin{center}
\includegraphics[width=0.6\columnwidth,keepaspectratio]{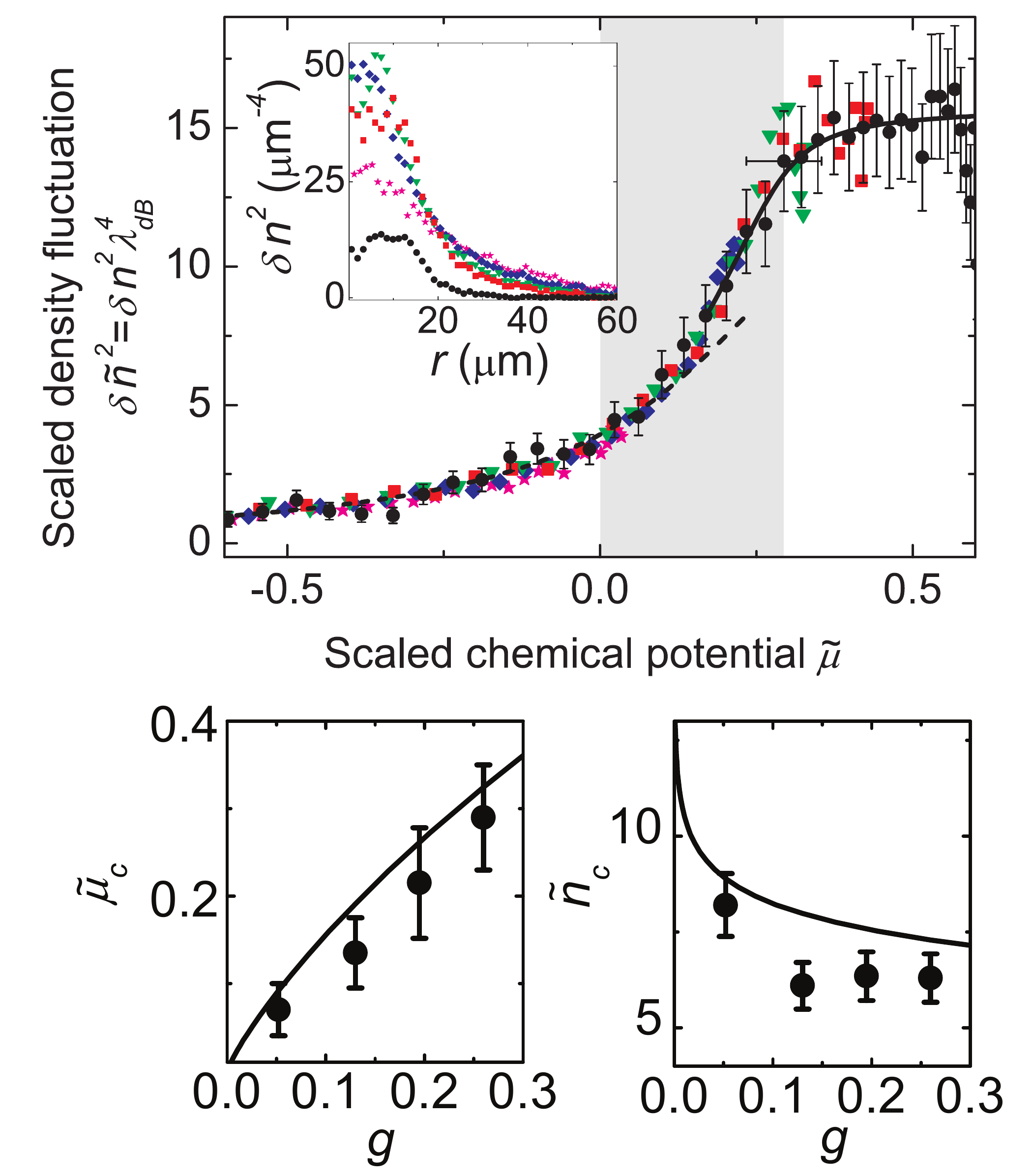}
\end{center}
\caption[Local density fluctuations of a 2D Bose gas.]{ Local density fluctuations and superfluid critical points of a 2D Bose gas. (Top panel) Local density fluctuations $\delta n^2 = \langle \delta N^2 \rangle/(A \lambda_\mathrm{dB}^2)$ are evaluated pixel-wise based on single-shot images taken under identical experimental conditions. After correcting for the systematics due to a finite point spread function discussed in Sec.~\ref{abs:mtf}, we extract the density fluctuation profile $\delta n^2(r)$ (inset), rescale and plot the fluctuation versus chemical potential according to the LDA discussed in Sec~\ref{eos}. Samples are prepared at five different temperatures: $T=21$~nK (black circles), 37~nK (red squares), 42~nK (green triangles), 49~nK (blue diamonds), and 60~nK (magenta stars), and a coupling constant $g=0.26$. Fluctuations in the normal gas phase are well described by the mean-field theory (dashed line). Solid line is an empirical fit to the crossover feature from which the critical chemical potential $\tilde{\mu}_c$ is determined. (Lower panels) Critical chemical potentials and critical phase space densities are determined from the crossover feature in the density fluctuation profiles and the corresponding value in the EoS measurements. Solid lines are theory predictions\cite{Prokof'ev01,Prokof'ev02}. Figures adapted from Hung et al.\cite{Hung11}} \label{fluctuation}
\end{figure}

As an example, the EoS $n(\mu,T)$ can be directly obtained from the density profile of a sample in equilibrium $n(\mathbf{r},T)$ using the substitution $\mu(r)+V(r)=\mu_0$, where $T$ and $\mu_0$ can be independently determined from fitting the low density regime of the sample with the known equation of state of a thermal gas, see Fig.~\ref{figEoS} (right panel) for the EoS of 2D gases and 2D lattice gases thus obtained.

\subsection{Local density fluctuations}\label{abs:fluctuations}
Another advantage of in situ imaging and the LDA is that each local subsystem can freely exchange particles with the rest of the trapped sample, which we can view as a particle reservoir. Free particle exchange with the reservoir introduces density fluctuations which are well described by the grand canonical ensemble. Measuring the local density fluctuations away from the mean value thus reveals equilibrium thermodynamic properties of the underlying many-body phases. The connection is established via the fluctuation-dissipation theorem (FDT)\cite{Huang63, Kubo66}, which links fluctuations to a system's response (susceptibility) to the coupled thermodynamic forces. In our case, local atom number $N$ is coupled with the local chemical potential $\mu$ and there is a direct correspondence between the local atom number fluctuations $ \delta N ^2 = \langle \delta N ^2 \rangle$ and the local compressibility $\kappa = \partial N/ \partial \mu$. In the thermodynamic limit, $ \delta N^2 =  k_B T \kappa $. The local density fluctuations are thus complementary to the equation of state, and are, in fact, very sensitive to the onset of phase transitions where a sudden change in the compressibility occurs.

\begin{figure}[h]
\begin{center}
\includegraphics[width=0.7\columnwidth,keepaspectratio]{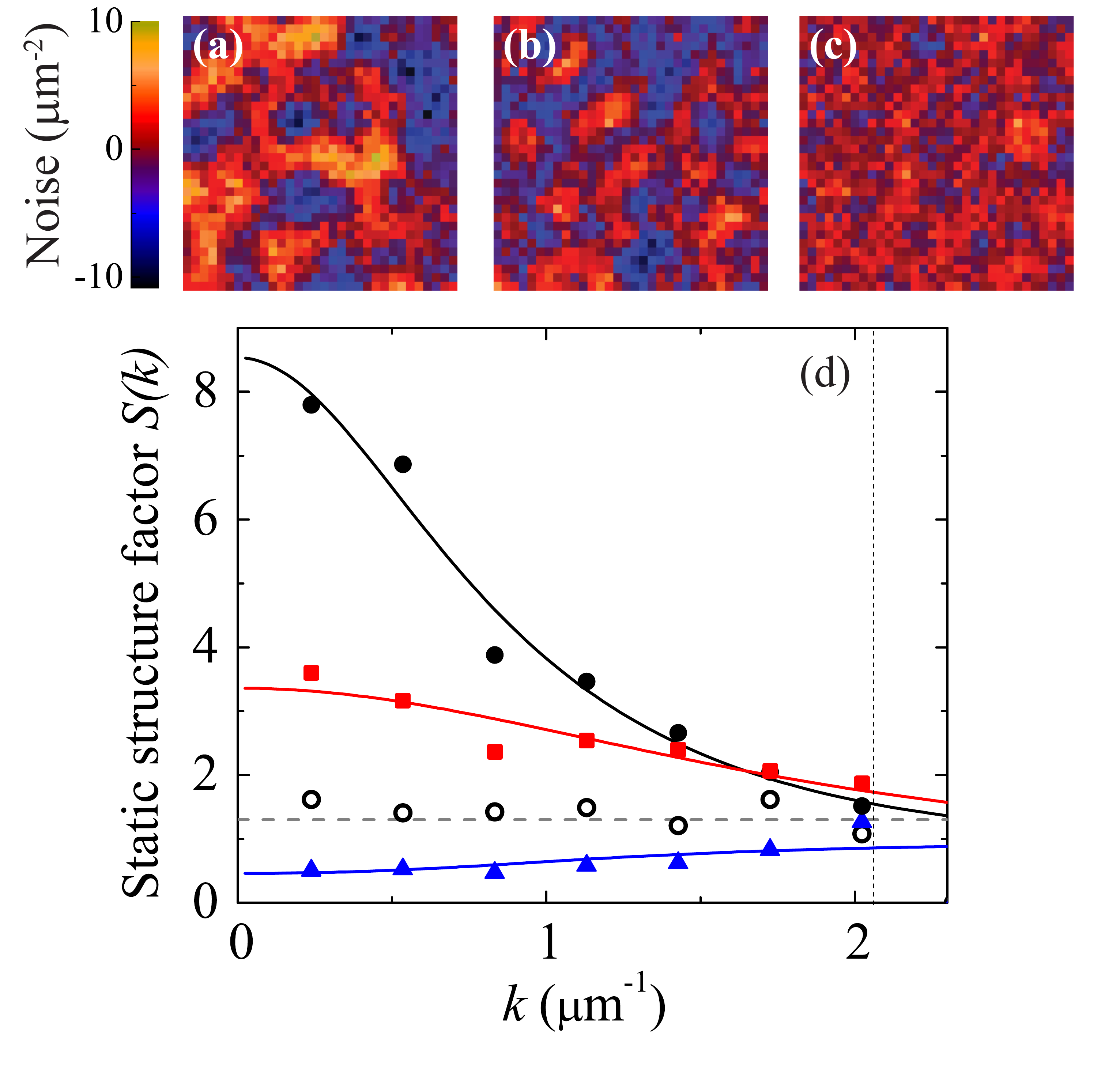}
\end{center}
\caption[Density fluctuations and static structure factors of 2D Bose gases.]{Density fluctuations and static structure factors of 2D Bose gases. Top panels show the sample noise images of weakly interacting 2D Bose gases (temperature $T=40~$nK) in the superfluid phase at dimensionless interaction strengths $g=$(a) $0.05$ and (b) 0.26, and (c) the noise image of a strongly interacting 2D lattice gas ($T=8~$nK) at an effective $g=1.0$. The static structure factor is extracted by calculating the density noise power spectrum, divided by the known image response function $\mathcal{M}^2(\mathbf{k})$ determined using high temperature, non-correlated 2D gases. Since the atomic interaction in the present sample is isotropic, and so is the correlation, the static structure factor $S(\mathbf{k})$ can be further averaged azimuthally and evaluated at points uniformly spaced in $k$, up to the resolution limited spatial frequency $k = 2\pi \mathrm{N.A.}/\lambda$. The results are shown in (d) for interacting 2D gases as shown in (a) (black circles), (b) (red squares), and (c) (blue triangles) together with the static structure factor of an ideal thermal gas at a phase space density $n\lambda^2_{dB}=0.5$ (open circles). The latter agrees with the theory value of $S(k) \approx 1.3$ for $k<\lambda_{dB}^{-1}=2~\mu$m$^{-1}$ (gray dashed line), providing a good validation of our analysis. Solid lines are the guides to the eye, generated based on the theory Eq.~\ref{Bogoluibov}. The vertical dashed line indicates the resolution limited spatial frequency $k=2\pi \mathrm{N.A.}/\lambda = 2.1~\mu$m$^{-1}$. Figures adapted from Hung et al.\cite{Hung11Corr}} \label{fig_sfactor}
\end{figure}

High-resolution in situ imaging is particularly suitable for detecting atomic density fluctuations. This is because the number of atoms within the resolution limited area is typically small ($N<$4) and hence, in a single-shot density image, the noise contribution from the local atom number fluctuation is comparable to or larger than the other technical noise such as optical shot noise of the imaging beam (see Fig.~\ref{fig_sfactor} a-c, for example).

Within a trapped sample, the local density fluctuations vary as a result of a spatially varying local chemical potential. One can similarly convert the measured fluctuation profile $ \delta n^2(\mathbf{r}) $ into $\delta n^2(\mu, T)$. As an example, in Fig.~\ref{fluctuation} we plot the density fluctuations of an interacting 2D quantum gas, where a sudden suppression of the density fluctuations occurs above a critical chemical potential. This feature can be associated with the onset of the superfluid phase transition in two dimensions. Meanwhile, the corresponding feature in the EoS shown in Fig.~\ref{figEoS} manifests only after we evaluate the first order derivative of the density with respect to the chemical potential, that is, the local compressibility $\kappa$\cite{Tung10}.

\subsection{Density-density correlation function and Static structure factor \label{abs:correlations}}
Beyond the local density fluctuations, more information regarding the collective excitations of an atomic quantum gas can be obtained from measuring the density correlations between two or more spatial points. Spatially resolved in situ measurements can be used to deduce these correlation functions. However, such measurements typically suffer from systematic errors caused by finite imaging resolution. This is because a finite image point spread function blurs the atomic signals into an area larger than or comparable to the real correlation length of the sample. This not only reduces the strength of the local fluctuation signal but also introduces artificial correlations. Here, we discuss a complementary, but highly effective correlation analysis in the Fourier domain (momentum space), where one can correct for the imaging systematics and deduce the sample's static structure factor, i.e. the Fourier transform of the sample's density-density correlation function.

We focus our discussion on two-point correlations of a 2D homogeneous sample. The density-density correlation depends on the separation $\mathbf{r}_1-\mathbf{r}_2$ between two points, and the static correlation function $\nu(\mathbf{r})$ is defined as \cite{Giogini98}
\begin{eqnarray}
\nu(\mathbf{r}_1-\mathbf{r}_2) &=&  \bar{n}^{-1}\langle \delta n(\mathbf{r}_1) \delta n (\mathbf{r}_2) \rangle \nonumber \\&=& \delta(\mathbf{r}_1-\mathbf{r}_2) + \bar{n}^{-1}\langle \hat{\psi}^\dag(\mathbf{r}_1) \hat{\psi}^\dag(\mathbf{r}_2) \hat{\psi}(\mathbf{r}_1)\hat{\psi}(\mathbf{r}_2)\rangle - \bar{n}\label{staticcorr},
\end{eqnarray}
where $\delta n(\mathbf{r}) = n(\mathbf{r})-\bar{n}$ is the local density deviation from its mean value $\bar{n}$. The Dirac delta function $\delta (\mathbf{r}_1-\mathbf{r}_2)$ represents the autocorrelation of individual atoms, and $\langle \hat{\psi}^\dag(\mathbf{r}_1) \hat{\psi}^\dag(\mathbf{r}_2) \hat{\psi}(\mathbf{r}_1)\hat{\psi}(\mathbf{r}_2)\rangle = G^{(2)}(\mathbf{r}_1-\mathbf{r}_2)$ is the second-order correlation function with $\hat{\psi}$ being the bosonic field operator \cite{Naraschewski99}.

The static structure factor, on the other hand, is the Fourier transform of the static correlation function \cite{Giogini98, Pitaevskii03}
\begin{eqnarray}
S(\mathbf{k}) &=&  \int \nu(\mathbf{r}) e^{- i \mathbf{k}\cdot\mathbf{r}} d^2 r,
\end{eqnarray}
where $\mathbf{k}$ is the spatial frequency wave vector. We can rewrite the static structure factor in terms of the density fluctuation in the reciprocal space as \cite{Pitaevskii03}
\begin{equation}
S(\mathbf{k}) = \frac{\langle \delta n(\mathbf{k}) \delta n(\mathbf{-k})\rangle}{N} = \frac{\langle |\delta n(\mathbf{k})|^2\rangle}{N},\label{s_powerspec}
\end{equation}
\noindent where $N$ is the total particle number. Here, $\delta n(\mathbf{-k}) = \delta n^*(\mathbf{k})$ since the density fluctuation $\delta n(\mathbf{r})$ is real. The static structure factor is therefore equal to the density fluctuation power spectrum, normalized to the total particle number $N$. An uncorrelated gas possesses a structureless, flat spectrum $S(\mathbf{k})=1$ while a correlated gas shows a non-trivial $S(\mathbf{k})$ for $k$ near or smaller than the inverse of the correlation length.

Using the analysis presented in Sec.~\ref{abs:mtf} to remove the contributions of imaging systematics and optical shot noise in the correlation measurements, we find that the static structure factor simply relates to the power spectrum of the measured density fluctuations as
\begin{equation}
 S(\mathbf{k}_l) \approx \frac{\langle |\delta n_{exp} (\mathbf{k}_l)|^2\rangle }{ N\mathcal{M}^2(\mathbf{k}_l)},\label{d_nexpmod}
\end{equation}
where, in the denominator, $\mathcal{M}^2(\mathbf{k})$ corrects for the response of the imaging optics at a given spatial frequency $\mathbf{k}$.
Note that the modulation transfer function $\mathcal{M}(\mathbf{k})$ has a natural cutoff at high $k$ due to the limited numerical aperture of the imaging optics. The static structure factor can thus only be extracted in a range of wavenumber $k$ that has non-zero $\mathcal{M}^2(\mathbf{k})$.

Following the above discussions, we show sample static structure factor measurements including a 2D thermal gas and interacting 2D Bose gases. To ensure that the extracted property reflects a homogeneous portion of the trapped sample, we limit the correlation analysis to a small area surrounding the trap center, where the variation of local chemical potential is minimal. Meanwhile, the area should also be large enough to offer sufficient resolution in Fourier space. Figure~\ref{fig_sfactor} shows typical density noise images used in the correlation analysis and the static structure factors evaluated from a series of such images.

The static structure factors of 2D thermal gases are routinely measured for calibration purposes. This is because their density-density correlations develop solely due to the bosonic bunching effect, and the static structure factor can be calculated from the known static correlation function of an ideal Bose gas\cite{Naraschewski99} $\nu(r)=\delta(r)+|g_1(z,e^{-\pi r^2/\lambda_{dB}^2})|^2/g_1(z,1)\lambda^2_{dB}$, where $z=e^{\mu/k_B T}$ is the local fugacity and $g_\gamma(x,y)=\sum_{k=1}^\infty x^ky^{1/k}/k^\gamma$ is the generalized Bose function. See Fig.~\ref{fig_sfactor} (d) for comparison between theory and experiment.

For weakly interacting 2D gases, the measurements also show consistency with the behavior predicted using the generalized fluctuation-dissipation theorem \cite{Pitaevskii03},
\begin{equation}
S(k) = \frac{\hbar^2 k^2}{2m \epsilon(k)} \coth \frac{\epsilon(k)}{2k_B T}\label{Bogoluibov},
\end{equation}
in which density correlations arise from the collective excitations that have a well-defined energy-momentum dispersion relation $\epsilon(k)$.

\section{Summary and outlook}
In this chapter, we have given a detailed account of the principle and experimental
implementation of in situ imaging atomic quantum gases in two dimensions. Particular
emphases have been given to the design of experimental setup for absorption imaging
and the optimization of imaging performance. From in situ images, we discuss two main
observables: density and fluctuations, and practical approaches of calibrating and removing
potential systematics.

We also offer detailed derivation of physical quantities that can be extracted from the
spatial distribution of density and fluctuations. Based on the local density approximation,
the density distributions of a sample in equilibrium can be converted into its equation of
state, while the fluctuations reveal the density-density structure factor. These quantities
provide a direct and essential information to test the scale invariance and universal properties of
2D atomic gases.

A number of new and intriguing research directions based on in situ imaging are currently been
explored at the time we are writing this chapter. One example is the non-equilibrium quantum dynamics
of atoms, which can be induced by deforming the trap \cite{Hung10} or Feshbach tuning of the interaction and recorded by monitoring the redistribution of the trapped atoms towards a new equilibrium \cite{Hung13}. Another interesting
application of the in situ imaging setup is to study quantum gases in arbitrary optical potentials,
e.g., box potential and exotic lattice configurations. In this case, a desired optical potential can
be formed by sending far-detuned light beams into the objective from the side of the eye piece
toward the atoms, and the higher resolution of the imaging system is, the smaller the feature size of
the potential one can produce.


\bibliographystyle{ws-rv-van}
\bibliography{chapterChin}

\end{document}